\documentclass{article}
\usepackage{amsmath}
\newcommand{\p}[1]{(\ref{#1})}
\newcommand{\cbZ}{\overline{\strut\cal Z}}
\newcommand{\cF}{{\cal F}}
\newcommand{\cZ}{{\cal Z}}
\newcommand{\cP}{{\cal P}}

\newcommand{\cA}{{\cal A}}
\newcommand{\cbF}{\overline{\strut\cal F}}
\newcommand{\bF}{{\overline{\rule{0pt}{0.8em} F}}}

\newcommand{\bX}{{\overline{\strut X}}{}}
\newcommand{\bY}{{\overline{\strut Y}}\,{}}
\newcommand{\bD}{{\overline{\strut D}}{}}
\newcommand{\bnabla}{{\overline{\strut\nabla}}{}}

\newcommand{\bxi}{{\bar\xi}}
\newcommand{\bpsi}{{\bar\psi}}

\newcommand{\be}{\begin{equation}}
\newcommand{\ee}{\end{equation}}
\newcommand{\bea}{\begin{eqnarray}}
\newcommand{\eea}{\end{eqnarray}}

\newcommand{\ba}{\begin{array}} \newcommand{\ea}{\end{array}}
\newcommand{\ds}{\displaystyle}

\renewcommand{\Im}{{\tt Im\,}}
\renewcommand{\Re}{{\tt Re\,}}
\newcommand{\sqrty}{\sqrt{\rule{0pt}{0.9em}y}}

\newcommand{\nn}{\nonumber}

\usepackage{amscd,amsmath,amssymb}
\usepackage{amsfonts}

\def\theequation{\arabic{section}.\arabic{equation}}

\begin{document}
\thispagestyle{empty}
\vspace{2cm}
\begin{flushright}
\end{flushright}\vspace{2cm}
\begin{center}
{\Large\bf N=4 Superconformal Mechanics and Black Holes}
\end{center}
\vspace{1cm}

\begin{center}
{\large\bf S.~Bellucci${}^{a}$, S.~Krivonos${}^{b}$,
A.~Shcherbakov${}^{a}$  and A.~Sutulin${}^{b}$}
\end{center}

\begin{center}
${}^a$ {\it
INFN-Laboratori Nazionali di Frascati,
Via E. Fermi 40, 00044 Frascati, Italy} \vspace{0.2cm}

${}^b$ {\it
Bogoliubov  Laboratory of Theoretical Physics, JINR,
141980 Dubna, Russia}
\end{center}
\vspace{2cm}

\begin{abstract}\noindent
The motion of a particle near the Reissner-Nordstr\"om black hole horizon is described by conformal mechanics. In this paper
we present an extended one-dimensional analysis of the~$N=4$ superconformal mechanics coupled to~$n$-copies of~$N=8$,~$d=1$ vector supermultiplets. The constructed system possesses a special K\"ahler geometry in the scalar
sector of the vector multiplets as well as an~$N=4$ superconformal symmetry which is provided by a proper coupling to a dilaton superfield. The superconformal symmetry completely fixes the resulting action. We explicitly demonstrate that
the electric and magnetic charges, presenting in the ``effective black hole'' action,  appear as a result of resolving
constraints on the auxiliary components of the vector supermultiplets. We present the component action, supercharges and Hamiltonian with
all fermionic terms included. One of the possible ways to generalize the black hole potential is to consider a modified version of the~$N=4$ superconformal multiplet where its auxiliary components acquire non-zero constant values. We explicitly write down the corresponding modified black hole potential.
\end{abstract}

\newpage
\setcounter{page}{1}

\section{Introduction}
Recently, the supersymmetric one dimensional theories appeared in the two different but related four-dimensional systems.
Firstly, in~\cite{bh1} it was established connection between black holes and conformal mechanics~\cite{AFF}.
Geodesic motion of a particle near the horizon of an extreme Reissner-Nordstr\"om black hole was shown to be related to the
relativistic version of the De~Alfaro--Fubini--Furlan conformal mechanics. Later on it was demonstrated~\cite{newcfm} that this
new version of relativistic conformal mechanics is just a known one rewritten in another coordinate system. Secondly, one-dimensional supersymmetric theories appear when considering a geodesic motion in the black hole background in the presence of~$N=2$,~$d=4$
vector supermultiplets~\cite{1,2}. The corresponding one dimensional action is fully specified by the metric of the scalars
which describes a  special K\"ahler geometry~\cite{KG}, and the effective black hole potential~$V_{BH}$, which depends only on
the scalars and electric and magnetic charges. The resulting system has been intensively studied also within attractor mechanism
framework~\cite{AM}.

In both these approaches the fermionic sectors are (almost) completely ignored. One reason for neglecting fermions is that most of needed features are really encoded in the bosonic sector of these theories. Another reason is that the dimensional reduction is not
straightforward, especially in the fermionic sector. That is why in the present paper we attempt to consider both these
one-dimensional theories with all fermionic terms included within one-dimensional supersymmetry framework.

The main topics we are going to consider include the following ones:
\begin{itemize}
\item We demonstrate how the effective black hole potential, with proper electric~$q_A$ and magnetic~$p^A$ charges, arises in a
theory with~$n$-copies of self-coupled~$N=8$,~$d=1$ vector supermultiplets (Section~\ref{sect2}). The important new peculiarities are the
appearance of the charges as a result of a specific solution  of the basic constraints on the auxiliary components
of the vector supermultiplets.

\item We analyze the structure of the action in terms of~$N=4$,~$d=1$ supermultiplets (Section~\ref{sect3}). This is a necessary step to
couple the vector supermultiplets to~$N=4$,~$d=1$ conformal supermultiplet. 
We split~$N=8$ vector supermultiplet in two~$N=4$ ones; the equivalence of these representations was first shown in~ \cite{iv}.
Each of these~$N=4$ supermultiplets possesses one physical boson and four fermions. The superfield constraints, describing the~$N=4$ supermultiplets, contain differential equation on two auxiliary components. Namely the solution to these equations
introduces~$p$-charges in the black hole potential. The~$q$-charges appear after dualization of the other auxiliary
components.

\item We consider the~$N=4$,~$d=1$ superconformal symmetry (Section~\ref{sectCONF}). The most important here is different conformal properties
of the two~$N=4$,~$d=1$ supermultiplets forming~$N=8$,~$d=1$ vector one. It turns out that with respect to the superconformal transformations, one of  them is a vector superfield, while the other is a scalar one. This completely fixes the action in terms
of~$N=4$ superfields. In the paper we deal  only with the case of~$SU(1,1|2)$ superconformal group.

\item Finally, we pass to the components and presented the general action with all fermionic terms included (Section~\ref{sectACTCOMP}).
The explicit form of the supercharges and the Hamiltonian is also given in this Section.

\item We also propose a possible generalization to the case where the superconformal multiplet itself contains non vanishing
coupling constant. The modified bosonic black hole potential is explicitly written (Section~\ref{sectMODIF}).
\end{itemize}

\setcounter{equation}0
\section{N=8 vector supermultiplet and BH potential}\label{sect2}
In this Section we demonstrate how the bosonic sector of the most general action for~$n$-vector~$N=8$,~$d=1$ supermultiplets
reproduces the main part of the BH action~\cite{{1},{2}} (with the supergravity sector being switched off). The constructed action
contains the effective ``black hole potential'' initially introduced in~\cite{2}. We also explain how the electric and magnetic charges,
entering this potential, appear within one-dimensional supersymmetry formalism.

The heart of the~$N=2$,~$d=4$ SYM theory is formed by a vector supermultiplet, which describes spin-1 particles, accompanied by
complex scalar fields and doublets of spinor fields. The geometry of the scalar fields is restricted to be a special
K\"ahler one~\cite{KG}. The restriction that the metric be defined by a holomorphic function is rather crucial for the consideration.

It is known for a long time that the standard reduction from the~$N=2$,~$d=4$ SYM to~$d=1$ gives
rise to a~$N=8$ supersymmetric theory with five bosonic fields, i.e., the {\bf (5,\;8,\;3)} supermultiplet\footnote{We use the notation
$\bf (n,N,N-n)$, in order to describe a supermultiplet with~$n$ physical bosons,~$N$ fermions and~$N-n$ auxiliary bosons.}~\cite{Z1,DE}.
Of course, after such a reduction the geometry of the bosonic sector is be completely different from the special K\"ahler one.
Fortunately, there is another version of the dimensional reduction from~$N=2$,~$d=4$ SYM to one dimension~\cite{bkn,bks1}.
This reduction is performed in terms of the field strength, rather than
in terms of a prepotential~\cite{Z1,DE}. In such an approach only a complex scalar of the vector multiplet becomes a physical boson in~$d=1$, while the rest of the bosonic components turns into auxiliary fields. Thus, we end up with the~{\bf (2,\;8,\;6)} supermultiplet.

The most general~$N=8$ supersymmetric action for~$n$-many vector supermultiplets can be written as an integral over~$N=4$ subspace of~${\mathbb R}^{1|8}$ superspace
\be\label{actionsf}
 S=- \int dt d^2\theta_2 d^2\vartheta_2\; \cF\left( \cZ{}^A \right) -
   \int dt d^2\theta_1 d^2\vartheta_1\; \cbF (\cbZ{}^A), \; \quad A=1, \ldots, n,
\ee
where~$\cF(\cZ{}^A)$ and~$\cbF(\cbZ{}^A)$ are arbitrary holomorphic functions of the vector supermultiplets~$\cZ{}^A$ and~$\cbZ{}^A$, respectively.
All relevant notations and definitions are given in appendix~A.

This action is notable for the fact that its bosonic part contains the well-known effective black hole potential~\cite{2}
\be\label{a3}
\ds S_{bos}=\int dt\left[ \rule{0pt}{1.2em} M_{AB} {\dot z}{}^A {\dot{\bar z}}{}^B - V(p,q,z,\bar z)\right],\quad
V=\frac1{16}\left( \begin{matrix} p & q \end{matrix} \right)
    \left( \begin{matrix} M + N M^{-1} N  & N M^{-1}\\
                          M^{-1} N & M^{-1}  \end{matrix} \right)
    \left( \begin{matrix} p \\ q \end{matrix} \right)
\ee
with the standard notations explained below, see formula~(\ref{mn}).
The appearance of the constants~$p^A$ and~$q_A$ is closely related to the proper treatment of the constraints on the auxiliary
components of the vector supermultiplet.
To clarify this point we integrate over the Grassmann variables in the action~(\ref{actionsf}) and take its bosonic part:
\be\label{a1}
S_{bos}= \int dt \left[ \left( F_{AB} +\bF_{AB}\right) {\dot z}{}^A {\dot{\bar z}}{}^B + \frac{1}{16}
\left( F_{AB}X^A Y^B +\bF_{AB}\bX^A \bY^B \right) \right].
\ee
The functions~$F_{AB}(z)$ and~$\bF_{AB}(\bar z)$ are nothing but a bosonic limit of the corresponding second derivatives of the prepotential~$\cF(\cZ)$ (see appendix~\ref{appendixA}).
The auxiliary components~$X^A$ and~$Y^A$ of the vector supermultiplet are not completely independent, but subjected to the  constraints~\p{constr1}:
\be\label{constr2}
\frac{\partial}{\partial t} \left( X^A-\bY^A \right) = 0,
\ee
while the rest four auxiliary bosonic components~$Y^{a\alpha}$ are expressed, as a consequence of their equations of motion, only in terms of the fermions and, therefore, they do not contribute to the bosonic part of the action.

To deal with the constraints~\p{constr2} let us firstly split them as follow
\bea
&& \frac{\partial}{\partial t}\, \mbox{Re}\left( X^A-Y^A \right) =0, \label{cc1} \\
&&\frac{\partial}{\partial t} \, \mbox{Im}\left( X^A+ Y^A \right) =0 . \label{cc2}
\eea
Then one may check that under the~$N=8$ supersymmetry,~$\mbox{Re }X^A$ and~$\mbox{Re }Y^A~$ transform as a total time derivative:
\be\label{t1}
\delta \mbox{Re }X^A = \delta \mbox{Re }Y^A  \sim \frac{\partial}{\partial t} \left( \mbox{fermions} \right).
\ee
So, we may solve~\p{cc1} as follows
\be\label{sol1}
\mbox{Re } X^A = \mbox{Re } Y^A = \partial_t \Upsilon{}^A
\ee
without breaking the~$N=8$ supersymmetry. Let us stress, that only due to the transformation properties~\p{t1} the right hand side
in~\p{sol1} can be represented as a time derivative of a new bosonic field~$\Upsilon{}^A$.

Concerning constraints~\p{cc2}, we just solve them as
\be\label{sol2}
\mbox{Im  }\left( X^A+ Y^A \right) =p^A = const,
\ee
where the constants~$p^A$ are identified as magnetic charges corresponding to the black hole~\cite{2}.
Substituting all these into the action~\p{a1} we get
\be\label{a2}
S_{bos} = \int dt \left[  M_{AB}\;{\dot z}{}^A {\dot{\bar z}}{}^B +\frac{1}{16} M_{AB}\left( {\dot\Upsilon}{}^A {\dot\Upsilon}{}^B -p^A p^B +c^A c^B\right)
 + \frac{1}{8}\;N_{AB}\;p^A{\dot\Upsilon}{}^B\right],
\ee
where
\be\label{mn}
M_{AB} = F_{AB}+\bF_{AB}, \quad N_{AB}=i\left( F_{AB}-\bF_{AB} \right), \qquad c^A=\mbox{Im }\left(X^A-Y^A\right).
\ee
Again, the auxiliary components~$c^A$ do not contribute to the bosonic part of the action and, consequently, might be neglected.

The coordinates~$\Upsilon^A$ are cyclic; the corresponding conserved quantities are identified with the electric charges~$q_A$:
\be\label{mom}
{\cal P}_{\Upsilon{}^A}=\frac{1}{8}\left(M_{AB}{\dot\Upsilon}{}^B+N_{AB}p^B\right) =-\frac18\,q_A=const.
\ee
Performing the Rauth transform with respect to the variable~$\Upsilon^A$ we get the action~(\ref{a3}) with~$V(p,q,z,\bar z)$ being exactly the effective black hole potential of Ref~\cite{2}. The full supersymmetric form of this action is given in the forthcoming Sections.

Despite the complete coincidence of the derived potential~$V(p,q,z,\bar z)$ with that one of Ref~\cite{2}, the actions do not coincide~-- it is easy to see that in the action~(\ref{a3}) it is missing one degree of freedom, namely that one corresponding to a dilaton. To make this field enter the game in a supersymmetric manner, one should introduce the interaction of the vector supermultiplet with an~$N=4$ superconformal  multiplet, which contains a dilaton among its components. Before doing this, let us first describe the vector supermultiplet in~$N=4$ setup, which is used to describe the superconformal multiplet.

\setcounter{equation}0
\section{The vector supermultiplet in $N=4$ superspace}\label{sect3}
The simplest way to introduce the interaction with the~$N=4$ conformal supermultiplet is to use~$N=4$ superfields.
So, let us discuss how the~$N=8$ vector supermultiplet is formulated in~$N=4$ superspace.

As a first step, we identify the~$SU(2)$ indices~$a$ and~$\alpha$ and introduce the following set of covariant derivatives
constructed from~$D^{ia},\nabla^{i\alpha }$~\p{sderiv}:
\be
\ba{l}
\ds D^a = \frac{1}{\sqrt{2}} \left( D^{1a} +i\nabla^{2a}\right), \; \bD{}^a=\frac{1}{\sqrt{2}} \left( D^{2a} +i\nabla^{1a}\right), \\[0.7em]
\ds \nabla^a = \frac{1}{\sqrt{2}} \left( D^{1a} -i\nabla^{2a}\right), \; \bnabla{}^a=\frac{1}{\sqrt{2}} \left( D^{2a} -i\nabla^{1a}\right).
\ea
\ee
Then we introduce  the real and imaginary parts of the vector multiplet~$\cZ, \cbZ$:
\be
U =\frac{1}{2}\left( \cZ+ \cbZ\right),\quad \Phi=-\frac{i}{2}\left( \cZ - \cbZ\right).
\ee
In terms of these variables the constraints (\ref{constr}a) have the following form
\be\label{n4con1}
D^a U+i\nabla^a \Phi=0, \; \nabla^a U+i D^a \Phi =0, \qquad \bD_a U -i \bnabla{}_a\Phi=0,\; \bnabla{}_aU-i\bD_a \Phi =0.
\ee
The role of these constraints is quite clear: they express $\nabla^a$ and~$\bnabla{}_a$ derivatives of the~$N=8$ superfields~$U$ and~$\Phi$
in terms of~$D^a$ and~$\bD_b$ ones. Therefore, all components of the vector supermultiplets are contained
in the~$N=4$ superfields~$u$ and $\phi$ depending on~$\theta_a-i\bar\vartheta_a$ and~$\bar\theta{}^a-i\vartheta^a$ only:
\be\label{uphi}
u=U\rule[-0.5em]{0.4pt}{1.6em}_{\,\vartheta=i\bar\theta,\bar\vartheta=i\theta}, \qquad
\phi=\Phi\rule[-0.5em]{0.4pt}{1.6em}_{\,\vartheta=i\bar\theta,\bar\vartheta=i\theta}.
\ee
The second set of the constraints (\ref{constr}b), being rewritten in the new basis, produces the irreducibility constraints on the
$N=4$ superfields~$u$ and~$\phi$:
\be\label{n4confin}
\left[D^a,\bD_a\right] u=0 \qquad (a)\;, \qquad \left( D^a\bD{}^b +D^b\bD{}^a \right) \phi =0 \qquad (b).
\ee
These constraints define the standard and twisted~$N=4$ supermultiplets, each containing one physical boson and four fermions~\cite{leva,ikl1}.
Let us note that from~(\ref{n4confin}a) it immediately follows that
\be\label{aab}
\frac{\partial}{\partial t} D^2 u = \frac{\partial}{\partial t} \bD{}^2 u =0,
\ee
where~$D^2=D^aD_a$ and~$\bD^2=\bD_a \bD^a$. These equations are just~$N=4$ variant of constraints~\p{constr1}.
Finally, using~\p{n4con1} one may rewrite the action~\p{actionsf} as an integral over the~$N=4$ superspace
\be\label{n4ac1}
 S=- \int dt d^2\theta d^2\bar\theta\;\left[  \cF\left( u^A+i\phi^A \right) +
   \cbF\left(u^A-i\phi^A\right)\right].
\ee
Last thing which has to be clarified is the appearance of the electric and magnetic charges within~$N=4$ formalism. It is rather easy to check that
the equation~\p{sol2} results in the conditions
\be\label{sol2a}
D^2 u^A=ip^A, \qquad \bD{}^2 u^A =-ip^A,
\ee
while the~$N=4$ analog of the condition~\p{sol1} reads
\be\label{sol1a}
i\left(D^2 \phi^A - \bD{}^2 \phi^A \right) = \partial_t \Upsilon{}^A.
\ee
Thus we have all the ingredients to reproduce the effective black hole action in terms of~$N=4$ superfields. In the next
Section we will consider~$N=4$ superconformal multiplet and will couple it to the black hole action.

To close this Section let us note that the equations~\p{sol2a} and~\p{sol1a} explicitly break the evident~$U(1)$ symmetry realized as
\be
\delta D^a = i\alpha D^a, \quad \delta \bD_a = -i\alpha \bD_a.
\ee
Instead of this symmetry another~$U(1)$ invariance appears in the system
\be\label{u1}
\delta D^a = \beta \bD{}^a, \quad \delta \bD_a = -\beta D_a.
\ee
It is  quite easy to check that the constraints~\p{n4confin},~\p{sol2a} and~\p{sol1a} are invariant with respect to this~$U(1)$ symmetry.
This fact will be important in the Section~\ref{sectACTCOMP} when introducing fermionic and auxiliary components with the proper charges with respect to this~$U(1)$ group.

\setcounter{equation}0
\section{Maintaining N=4 superconformal symmetry}\label{sectCONF}
In this Section we are going to couple the effective black hole action~\p{n4ac1} to the~$N=4$ superconformal multiplet.
In one dimension the most general superconformal group is~$D(2,1;\alpha)$ one~\cite{sorba}. Here we restrict
our consideration by the special case with~$\alpha=-1$ which corresponds to~$SU(1,1|2)$ symmetry. This superconformal
group has natural realization in the~$N=4$,~$d=1$ superspace~\cite{leva}
\be\label{sc1}
\delta t = E -\frac{1}{2} \theta^a D_a E -\frac{1}{2}\bar\theta_a \bD{}^a E, \quad \delta \theta^a=-\frac{i}{2} \bD{}^a E,\quad
\delta \bar\theta=-\frac{i}{2} D_a E,
\ee
where the superfunction~$E(t,\theta,\bar\theta)$ collects all~$SU(1,1|2)$ parameters:
\be\label{E}
E=f(t)-2i \left( \varepsilon \bar\theta -\theta\bar\varepsilon \right) + \theta^a \bar\theta{}^b B_{(ab)}+
  2\left( \dot\varepsilon \bar\theta +\theta\dot{\bar\varepsilon}\right)(\theta\bar\theta) +\frac{1}{2} (\theta\bar\theta)^2
   \ddot{f}.
\ee
Here
\be
f=a+bt+ct^2,\quad \varepsilon^a = \epsilon^a + t \eta^a.
\ee
The bosonic parameters~$a,b,c$ and~$B_{(ab)}$ correspond to translations, dilatations, conformal boosts and rigid~$SU(2)$ rotations,
while fermionic parameters~$\epsilon^a$ and~$\eta^a$ correspond to Poincar\'e and conformal supersymmetries, respectively.

It is important that by definition the function~$E$ obey conditions
\be
D^2 E = \bD{}^2E= \left[D^a,\bD_a\right]E=0, \qquad \partial^3_t E = \partial^2_t D^a E =\partial_t D^{(a}\bD{}^{b)}E=0.
\ee
As usually, we consider the systems where the conformal supersymmetry is spontaneously broken. Therefore, to maintain~$SU(1,1|2)$ invariance, one has to introduce the super-dilaton~---~$N=4$,~$d=1$ superfield~$\tilde u$ which transforms as follows
\be\label{dilaton}
\delta {\tilde u} = \partial_t E.
\ee
It is known for a long time~\cite{leva} that the super-dilaton has to be further constrained by the conditions
\be\label{y}
D^2 Y = \bD{}^2 Y = \left[D^a, \bD_a\right] Y=0, \qquad Y= e^{\tilde u}.
\ee
These constraints are fully compatible with the transformation properties~\p{sc1},~\p{dilaton}.

The invariant superspace measure is defined as
\be\label{sc2}
\triangle s = dt d^2 \theta d^2 \bar\theta \; Y,
\ee
while the invariant action for superconformal multiplet~$Y$ is unambiguously restored to be
\be\label{sc3}
S_g = -\int dt d^2 \theta d^2 \bar\theta \; Y \log{Y}.
\ee
Clearly enough, the invariant action for the matter fields~$\cA$ can be constructed as
\be\label{matter1}
S_{matter}=-\int dt d^2 \theta d^2 \bar\theta \; Y\; G(\cA),
\ee
provided the superfields~$\cA$ are scalars under the superconformal transformations. In black hole action~\p{n4ac1} we have two
sets of the~$N=4$ superfields:~$u^A$ and~$\phi^A$ restricted by the constraints~\p{n4confin}.  Keeping in the mind the transformation
properties of the covariant spinor derivatives~$D^a, \bD_a$ under~$SU(1,1|2)$
\be
\delta D^a =-\frac{i}{2} \left( D^a \bD_b E\right)D^b, \qquad \delta \bD_a =-\frac{i}{2}\left( \bD_a D^b E\right) \bD_b
\ee
one may check that the constraints~\p{n4confin} are invariant under the~$N=4$ superconformal group if the superfields~$u^A, \phi^A$
have the following transformation properties
\be\label{sc4}
\delta u^A = \partial_t E\; u^A, \qquad \delta \phi^A=0.
\ee
So, the superfields~$\phi^A$ are scalars under superconformal transformations, while~$u^A$ are vectors. The simplest way to construct
scalars from the superfields~$u^A$ is to consider the combinations~$(u^A Y^{-1})$. Therefore, the superconformally invariant
black hole action reads
\be\label{Action}
S=-\int dt d^2 \theta d^2 \bar\theta \;Y\;\left[  \log{Y}+ \cF\left( \frac{u^A}{Y}+i\phi^A \right) +
   \cbF\left(\frac{u^A}{Y}-i\phi^A\right)\right],
\ee
where the~$N=4$ superfields are constrained by the conditions~\p{n4confin},\p{sol2a} and~\p{y}.

In the next Section we will consider this action on the component level and with the fields~$\Upsilon^A$ introduced and then
dualized into the charges~$q_A$.

It is worth mentioning that there is another possibility to construct scalar superfields from~$u^A$ ones. Indeed, the ratio of any two
superfields~$u^A$, for example the superfields~$u^A/u^1$, are perfectly scalars with respect to the superconformal transformations. So, an arbitrary
function of these superfields is a good candidate to be an invariant Lagrangian density. The full analysis of a such situation is out
of the scope of the present paper and will be considered elsewhere.

\setcounter{equation}0
\section{Field content of the supersymmetric black hole action}\label{sectACTCOMP}
To find the components action one has to integrate over the Grassmann variables in the superfield action~\p{Action}, remove the auxiliary components
through their equations of motion and perform the dualization in a way discussed in the previous sections. Before presenting the component
action let us define the physical bosonic and fermionic components as
\be
\ba{lll}
\ds z^A=\frac{u^A}Y + i \phi^A, & \ds \quad  {\bar z}^A = \frac{u^A}Y - i \phi^A, &\ds \quad y = Y\\
\ds \eta^{aA} = \sqrt{\frac{Y}{2}} (iD^a -\bD^a) z^{A}, &\ds \quad \chi^{aA} = \sqrt{\frac{Y}{2}} (iD^a - \bD^a)\bar z^A,&
    \ds \quad \rho^a = \frac{1}{\sqrt{\strut 2Y}}\, (iD^a -\bD^a) Y, \\
\ds \bar \eta_a^A = \sqrt{\frac{Y}{2}} (i\bD_a - D_a) \bar z^A , &\ds \quad \bar \chi_a^A =\sqrt{\frac{Y}{2}} (i \bD_a -D_a) z^A ,&
    \ds \quad {\bar\rho}_a = \frac{1}{\sqrt{\strut 2Y}}\, (i\bD_a - D_a) Y,
\ea
\ee
where in the right hand sides it is supposed to be taken only the first components in their Grassmann decompositions.
Being integrated over the Grassmann variables, the expression~(\ref{Action}) acquires the following form
\be\label{fa}
\ba{l}
\ds S = \int dt \left[
\frac1y\, {\dot y}{}^2 + y M_{AB} {\dot z}{}^A \dot{\bar z}{}^B - \frac1y\, V(p,q,z,\bar z)
\right.\\
\ds\phantom{\ds S = \int dt \left[ \right]}
\left.
    + i \left(\dot\rho^a \bar\rho_a -\rho^a\dot{\bar\rho}_a\right)
    +\frac{i}2\, M_{AB}\,\left( \dot\eta^{aA} \bar\eta^B_a -\eta^{aA}\dot{\bar\eta}_a^B
    + \dot\chi^{aA} \bar\chi_a^B  - \chi^{aA}\dot{\bar\chi}_a^B\right)
\right.\\
\ds\phantom{\ds S = \int dt \left[ \right]}
\left.
    + \frac{i}2\,M_{AB} {\dot z}^A \left(\rho^a \bar\eta^B_a - \chi^{aB}\bar\rho_a\right)
    + \frac{i}2\, M_{AB} {\dot{\bar z}}{}^A\left(\rho^a \bar\chi^B_a - \eta^{aB} \bar\rho_a \right)
\right.\\
\ds\phantom{\ds S = \int dt \left[ \right]}
\left.
    +\frac{i}{2}\left(F_{ABC}{\dot z}^A - \bF_{ABC}\dot{\bar z}{}^A\right) \left( \eta^{aB} \bar\eta^C_a-\chi^{aB}\bar\chi^C_a\right)
\right.\\
\ds\phantom{\ds S = \int dt \left[ \right]}
\left.
    - \frac i{8y} \left(q_A - 2 i \bF_{AB}p^B \right)\left( \rho^a\bar\chi^A_a + \eta^{aA}\bar\rho_a + 2 M^{AE}F_{CDE}\eta^{aC}\bar\chi^D_a \right)
\right.\\[0.8em]
\ds\phantom{\ds S = \int dt \left[ \right]}
\left.
    + \frac i{8y} \left(q_A + 2 i F_{AB}p^B \right)\left( \rho^a\bar\eta^A_a + \chi^{aA}\bar\rho_a + 2 M^{AE}\bF_{CDE}\chi^{aC}\bar\eta^D_a \right)
 \right]+\ldots
\ea
\ee
The expression for~$V(p,q,z,\bar z)$ is given earlier by the formula~(\ref{a3}). The dots stand for four-fermionic terms; they do not depend on the charges~$p^A$,~$q_A$ and since their explicit form is not too
illuminated, they are written down only in the Hamiltonian.

The explicit form of the on-shell action~\p{fa} together with the~$N=4$ supersymmetry transformations~\p{dilaton},~\p{sc4} provides all ingredients needed to construct the supercharges and Hamiltonian. As usual, the structure of the fermionic momenta
means that the system possesses second class constraints and, therefore, one has pass to Dirac brackets. The rest of the calculations goes straightforwardly, so we omit all details and present the final results.

The non-vanishing Dirac brackets between the canonical variables read
\bea\label{PB}
&& \left\{y, \cP_y\right\}=1, \; \left\{z^A, \cP_B\right\}=\delta^A_B,\; \left\{ {\bar z}{}^A,{\bar\cP}_B\right\}=\delta^A_B,\;
\left\{ \cP_A ,{\bar\cP}_B\right\}=i M^{CD}F_{ACE}\bF_{BDF}\left(\eta^{aE}{\bar\eta}_a^F-\chi^{aF}\bar\chi_a^E\right), \nn\\
&&\left\{ \cP_A ,\eta^{aB} \right\}=M^{BC}F_{ACD}\eta^{aD},\; \left\{ \cP_A ,\bar\chi{}^{aB} \right\}=M^{BC}F_{ACD}\bar\chi{}^{aD},\nn\\
&&\left\{ {\bar\cP}_A ,\bar\eta{}^{aB} \right\}=M^{BC}\bF_{ACD}\bar\eta{}^{aD},\;
\left\{ {\bar\cP}_A ,\chi^{aB} \right\}=M^{BC}\bF_{ACD}\chi^{aD},\nn\\
&& \left\{ \rho^a,\bar\rho_b\right\}=\frac i2\,\delta^a_b, \; \left\{ \eta^{aA},\bar\eta{}^B_b\right\}=i M^{AB}\delta^a_b,\;
\left\{ \chi^{aA},\bar\chi{}^B_b\right\}=i M^{AB}\delta^a_b.
\eea
The supercharges can be represented as
\bea\label{SC1}
&& {\mathbb Q}^a = Q^a + \frac1{\sqrty}\left[ \rule{0pt}{1.2em} {\cal Q}^a - \bar{\cal S}^a \right]
    -\frac{i}{\sqrty}\, M_{AB} \left[\eta^{(aA}\bar\eta{}^{b)B}+\chi^{(aA}\bar\chi{}^{b)B} \right]\rho_b,\nn\\
&&\bar{\mathbb Q}_a = {\bar Q}_a + \frac1{\sqrty} \left[\rule{0pt}{1.2em} \bar{\cal Q}_a+{\cal S}_a \right]
    +\frac{i}{\sqrty}\, M_{AB}\left[\eta^{A}_{(a}\bar\eta{}^B_{b)}+ \chi^{A}_{(a}\bar\chi{}^{B}_{b)}\right]\bar\rho^b.
\eea
Here we split up the supercharges to make its structure clearer: it contains pure dilaton contribution~$Q^a$, a pair of supercharges~${\cal Q}^a$ and~${\cal S}^a$, corresponding to the vector supermultiplet, and an ``interference'' term:
\be\label{QQQ1}
\ba{l}
\ds Q^a = \sqrty \rho^a \cP_y +\frac{i}{2\sqrty}\,\rho^2 \bar\rho^a, \\
\ds {\cal Q}^a =\eta^{aA}\cP_A-\frac i2\, {\bF}_{ABC}\chi^{bA}\chi^B_b\bar\eta^{aC} + \frac14 \left( q_A - 2i F_{AB} p^B \right)\bar\eta^{aA},   \\[0.5em]
\ds {\cal S}^a = \bar\chi^{aA}\cP_A + \frac i2 {\bF}_{ABC} \bar\eta^A_b\bar\eta^{bB}\chi^{aC} - \frac14 \left( q_A - 2i {\bF}_{AB} p^B \right)\chi^{aA}. \ea
\ee
The supercharges~$\mathbb Q$ form~$N=4$ superalgebra
\be
\left\{ {\mathbb Q}^a ,\bar{\mathbb Q}_b\right\}=2i\delta^a_b {\mathbb H}, \qquad
\left\{ {\mathbb Q}^a ,{\mathbb Q}_b\right\}= \left\{ \bar{\mathbb Q}^a ,\bar{\mathbb Q}_b\right\}
    = \left\{ {\mathbb H}, {\mathbb Q}_b\right\} = \left\{ {\mathbb H}, \bar {\mathbb Q}_b\right\}=0,
\ee
where the Hamiltonian reads
\be
\ba{l}
\ds {\mathbb H}=\frac1y\,M^{AB} \cP_A \bar\cP_B + \frac 1y\, V + \frac14\, y \cP_y^2 + \frac1{8y}\, \rho^2\bar\rho^2\\
\ds \phantom{{\mathbb H}=} -\frac i{2y}\, \cP_A \left( \rho^a \bar\chi_a^A+\bar\rho_a \eta^{aA}\right)
    -\frac i{2y}\, \bar\cP_A \left( \rho^a \bar\eta_a^A+\bar\rho_a \chi^{aA}\right)\\[0.8em]
\ds \phantom{{\mathbb H}=}
    + \frac i{8y}\, M^{AD} F_{ABC}\left( \eta^{aB}\eta_a^C + \bar\chi_a^B \bar\chi^{aC} \right)
    \left( q_D - 2i \bF_{DE} p^E\right)\\[0.9em]
\ds \phantom{{\mathbb H}=}
     - \frac i{8y}\, M^{AD} \bF_{ABC}\left( \bar\eta_a^B \bar\eta^{aC}+ \chi^{aB}\chi_a^C \right)
    \left( q_D + 2i F_{DE} p^E\right)\\[0.8em]
\ds \phantom{{\mathbb H}=}
    + \frac i{8y}\,q_A \rho^a \left( \eta^A_a - \chi^A_a\right)
    - \frac i{8y}\,q_A \bar\rho_a \left(\bar\eta^{aA} - \chi^{aA}\right)
    + \frac 1{2y}M_{AB} \rho^{(a}\bar\rho^{b)} \left( \eta_{(a}^A \bar\eta_{b)}^B + \chi_{(a}^A \bar\chi_{b)}^B \right)\\[0.8em]
\ds \phantom{{\mathbb H}=}
    - \frac 1{4y}\, p^A F_{AB} \left(\rho^a \eta_a^A + \bar\rho_a \bar\chi^{aA}\right)
    - \frac 1{4y}\, p^A \bF_{AB} \left(\rho^a \chi_a^A + \bar\rho_a \bar\eta^{aA}\right)\\[0.8em]
\ds \phantom{{\mathbb H}=}
    - \frac1y\, M^{AD} F_{ABC} \bF_{DEF} \eta^{aB} \eta_a^C \chi^{aE} \bar\chi_a^F
    - \frac 1{4y}\, M_{AB} M_{CD} \eta^{(a A}\bar\eta^{b)B} \chi_{(a}^C \bar\chi_{b)}^D \\[0.8em]
\ds \phantom{{\mathbb H}=}
    - \frac1{4y}\, F_{ABC} \left( \eta^{aA}\eta_a^B \bar\rho_b \bar\chi^{bC} + \bar\chi_a^A\bar\chi^{aB} \rho^b \chi_b^C\right)
    - \frac1{4y}\, \bF_{ABC} \left( \chi^{aA}\chi_a^B \bar\rho_b \bar\eta^{bC} + \bar\eta_a^A\bar\eta^{aB} \rho^b \chi_b^C\right)\\[0.8em]
\ds \phantom{{\mathbb H}=}
    + \frac1{4y}\, \left( \bF_{ABCD} - M^{EF} \bF_{ABE} \bF_{CDF} - 2M^{EF}\bF_{ACE} \bF_{BDF}\right)
        \chi^{aA}\chi_a^B \bar\eta^C_b\bar\eta^{bD}\\[0.8em]
\ds \phantom{{\mathbb H}=}
    + \frac1{4y}\, \left( F_{ABCD} - M^{EF} F_{ABE} F_{CDF} - 2M^{EF}F_{ACE} F_{BDF}\right)
        \eta^{aA}\eta_a^B \bar\chi^C_b\bar\chi^{bD}\\[0.8em]
\ds \phantom{{\mathbb H}=}
    +\frac1{4y}\left( M^{AD} F_{ABC} \bF_{DEF} - \frac12\, M_{BC} M_{EF}  \right)
    \left( \eta^{(aB}\bar\eta^{b)C} \eta_{(a}^E \bar\eta_{b)}^F + \chi^{(aB}\bar\chi^{b)C} \chi_{(a}^E \bar\chi_{b)}^F \right).
\ea
\ee
The highlighted structure of the supercharges~${\mathbb Q}^a,\bar{\mathbb Q}_a$ is not accidental. One may check that each
set of the sub-supercharges~$Q^a,{\bar Q}_b$,~${\cal Q}^a,\bar{\cal Q}_b$ and~${\cal S}^a,\bar{\cal S}_b$ independently forms~$N=4$ superalgebra
\be
\left\{Q^a,{\bar Q}_b\right\}=2i\delta^a_b H, \quad
\left\{{\cal Q}^a,{\bar{\cal Q}}_b\right\}=2i\delta^a_b{\cal  H}, \quad
\left\{{\cal S}^a,{\bar{\cal S}}_b\right\}=2i\delta^a_b {\cal  H},
\ee
where the Hamiltonians of the conformal and vector multiplets have the form:
\be
\ba{l}
\ds H= \frac14\, y \cP_y^2 + \frac1{8y}\, \rho^2\bar\rho^2, \\
\ds {\cal H}= \frac12 M^{AB} \cP_A \bar{\cP}_B
    + \frac12\, V \\
\ds \phantom{H=}
    + \frac i{16}\, M^{AD} F_{ABC}\left( \eta^{aB}\eta_a^C + \bar\chi_a^B \bar\chi^{aC} \right)
    \left( q_D - 2i \bF_{DE} p^E\right)\\[0.5em]
\ds \phantom{H=}
     - \frac i{16}\, M^{AD} \bF_{ABC}\left( \bar\eta_a^B \bar\eta^{aC}+ \chi^{aB}\chi_a^C \right)
    \left( q_D + 2i F_{DE} p^E\right)\\[0.5em]
\ds \phantom{H=}+ \frac18\, \left( \bF_{ABCD} - M^{EE'} \bF_{ABE} \bF_{CDE'} - 2M^{EE'}\bF_{ACE} \bF_{BDE'}\right)
        \chi^{aA}\chi_a^B \bar\eta^C_b\bar\eta^{bD}\\[0.5em]
\ds \phantom{H=}+ \frac18\, \left( F_{ABCD} - M^{EE'} F_{ABE} F_{CDE'} - 2M^{EE'}F_{ACE} F_{BDE'}\right)
        \eta^{aA}\eta_a^B \bar\chi^C_b\bar\chi^{bD}\\[0.5em]
\ds \phantom{H=}+ \frac18\, M^{AD} F_{ABC} \bF_{DEF}
    \left(\strut\chi^{aB}\chi_a^C \bar\eta^E_b\bar\eta^{bF} + \eta^{aB}\eta_a^C \bar\chi^E_b\bar\chi^{bF}
    - 4 \eta^{aB}\bar\eta_a^C \chi^E_b\bar\chi^{bF}
    \right).
\ea
\ee
As it was mentioned above, the supercharges~$Q^a,{\bar Q}_b$ are recognized as the supercharges of one dimensional~$N=4$ super\-conformal
mechanics~\cite{leva,ikp}, while the mutually anticommuting supercharges~${\cal Q}^a$ and~${\cal S}^a$ span~$N=8$ superalgebra with vanishing central charge~\cite{bkn,bks1}. Thus, coupling of the vector supermultiplet to the superconformal one goes through
the ``diagonal'' centerless~$N=4$ superalgebra spanned by the combinations~${\cal Q}^a-\bar{\cal S}^a$ in the full~$N=8$ supersymmetry
existing in the case on~$n$-copies of~$N=8$,~$d=1$ vector supermultiplets.

\setcounter{equation}0
\section{Black hole potential modification}\label{sectMODIF}
{}From the point of view of the superconformal group~$SU(1,1|2)$, the superconformal multiplet is defined as an exponential of the
dilaton superfield~$\tilde u$. The covariance with respect to the superconformal transformations fixes the irreducibility constraints to
be~\p{y}. Really speaking, these constraints may be slightly modified as
\be\label{Y}
D^a D_a Y = i m, \qquad \bD_a \bD^a Y = - i m, \qquad [D^a, \bD_a] Y = 0.
\ee
This seems to be the only possible modification of the constraints supporting superconformal symmetry and~$U(1)$ symmetry~\p{u1}.

The supersymmetric action preserves its previous form, but due to the presence of the constant~$m$, its component form gets modified
(we consider here only its bosonic part)
\be
\ba{l}
\ds S=\int dt \left[ y M_{AB} \dot z{}^A \dot{\bar z}{}^B
    + \frac{\dot y^2}y
    - \frac{m^2}{4y}
    - \frac m4 (Y^A-\bY^A)(F_A - {\bF}_A)
    +  y M_{AB} c^A c^B\right.\\
\ds\phantom{S=\int dt}-\left.\frac14\,  y F_{AB} \left( \frac{X^A-i m\; \Re z^A}y + i Y^A\right)\left( \frac{\bX^B + i m\; \Re z^B}y
    + i \bY^B\right)\right. \\[0.3em]
\ds\phantom{S=\int dt}-\left. \frac14\,y {\bF}_{AB} \left( \frac{X^A-i m\; \Re z^A}y - i Y^A\right)\left( \frac{\bX^B + i m\; \Re z^B}y
    - i \bY^B\right) \right].
\ea
\ee
It is obvious that on-shell in the pure bosonic limit the auxiliary fields~$c^A$ vanish. The dualization goes the same way as it
is described in the previous section: one should declare the auxiliary field~$X^A$ be a constant (again, due to~$U(1)$ arguing, it has to
be an imaginary one), while the field~$Y^A$ be split on the real and imaginary parts
$$X^A = i p^A,\quad \bX^A = - i  p^A, \quad Y^A=\Re Y^A + i \Im Y^A.$$
The later field has to be dualized~$\Im Y^A=\partial_t \Upsilon^A$, while the former~-- eliminated as it usually happens to auxiliary
fields. The result is
\be
S=\int dt \left[ y M_{AB} \dot z{}^A \dot{\bar z}{}^B + \frac{\dot y^2}y -\frac1y V_{eff} \right], \qquad
V_{eff} = \frac{m^2}{4y} + V(\tilde p,\tilde q,z,{\bar z})\\
\ee
The form of the black hole potential~(\ref{a3}) remains unchanged apart from the fact that it acquires dependence on new ``charges''
$$\tilde p^A=p^A - m\, \Re z^A, \qquad \tilde q_A = q_A - i m \left( F_A - {\bF}_A\right).$$
Thus we see that this modification results in the field-dependent shifts of the electric and magnetic charges~$q_A$ and~$p^A$.
The detailed discussion of such a modification will be considered elsewhere.

\section*{Conclusion}
In this paper we analyzed a system constructed by coupling of~$n$-copies of~$N=8$,~$d=1$ vector supermultiplets to~$N=4$ superconformal one. The~$N=4$ superconformal symmetry uniquely fixes the resulting action. We demonstrated that
the electric and magnetic charges, presenting in the ``effective black hole'' action~\cite{2} appear as a result of solving the
constraints on the auxiliary components of the vector supermultiplets. We also present the full component action with
all fermionic terms included as well as the supercharges and Hamiltonian. The one of the possible ways to generalize the
black hole potential is to consider modification of the~$N=4$ superconformal multiplet by assigning  non vanishing constant values to its auxiliary fields. We explicitly wrote down the corresponding modified black hole potential.

In some sense this paper can be considered as a continuation of the studies initiated in~\cite{bh1}. Keeping in mind that
the ``new conformal mechanics'' constructed in~\cite{bh1} is nothing but standard one~\cite{AFF} rewritten in terms of other
coordinates~\cite{newcfm}, one may hope that the constructed system describes the dynamics of a particle near the horizon
of an extreme Reissner-Nordstr\"om black hole when the interaction with electromagnetic fields is turned on. The dimensional reduction from~$d=4$ to~$d=1$ is not straightforward in this case. Therefore, the fully one-dimensional
consideration we presented in this paper could be useful.

Another very interesting area of applications of the system considered is a possible relation to a one-dimensional
geodesic motion in a black hole background in the presence of~$N=2, d=4$ vector supermultiplets~\cite{2}. The main problem
in this direction is that our bosonic action is different from those one in~\cite{2} by a factor. Of course,
this difference appears because we insist on the~$N=4$ superconformal symmetry which is absent in the action of~\cite{2}.
Nevertheless, it sounds reasonable that our action could appear in some limit (near BH horizon?). If it happens,
the full action with all fermionic terms included could help to analyze the attractor mechanism, etc. Related question
is whether it is possible to keep all~$N=8$ supersymmetry unbroken? To construct such an action one has to introduce the interaction with~$N=8$ superconformal multiplet. Unfortunately, up to now such a multiplet, which admits a non-trivial
potential term in the action, is known to have only on-shell description~\cite{leva}. But this is not the problem provided we are working in the Hamiltonian formalism. We are planning to consider
such a system elsewhere.

\section*{Acknowledgements}

We are grateful to Sergio~Ferrara, Alessio~Marrani  and Armen~Yeranyan for clarifying discussions.
\vskip .2cm
S.K. and A.Su. thank Laboratori Nazionali di Frascati, INFN, where this work was completed for kind hospitality
and financial support.  This work was partially supported by the European Community Human Potential
Program under contract MRTN-CT-2004-005104 \textit{``Constituents, fundamental forces and symmetries of the universe''},
by  INTAS under contract 05-7928 and by grants RFBR-06-02-16684, 06-01-00627-a and  DFG~436 Rus~113/669/03.

\def\theequation{\Alph{section}.\arabic{equation}}
\setcounter{equation}{0}
\setcounter{section}{1}
\appendix
\section{Appendix}\label{appendixA}

The natural framework to describe~$N=8$ vector supermultiplet is the~$N=8$,~$d=1$ superspace~$\mathbb{R}^{1|8}$
$$
\mathbb{R}^{1|8}=(t,\theta^{ia},\vartheta^{i\alpha})\,,\qquad
\left(\theta^{ia}\right)^\dagger=\theta_{ia}\,,\qquad
\left(\vartheta^{i\alpha}\right)^\dagger=\vartheta_{i\alpha}\,,$$
where~$i,\,a,\,\alpha=1,\,2$ are doublet indices of three~$SU(2)$ subgroups of
the automorphism group of~$N=8$ superspace.~$SU(2)$ metric is given by the skew-symmetric tensor
$$\epsilon_{ij} \epsilon^{jk}=\delta_i^k,\qquad \epsilon_{12} = \epsilon^{21} =1.$$

In superspace~$\mathbb{R}^{1|8}$ we define covariant spinor derivatives satisfying the following superalgebra
\be\label{sderiv}
\left\{D^{ia},D^{jb}\right\}=2i\epsilon^{ij}\epsilon^{ab}\partial_t\,,\quad
\left\{\nabla^{i\alpha},\nabla^{j\beta}\right\}=
2i\epsilon^{ij}\epsilon^{\alpha\beta}\partial_t\,.
\ee
In full analogy with~$N=2, d=4$ SYM, we introduce
a complex scalar~$N=8$ superfield~$\cZ$ subjected to the following constraints:
\be \label{constr}
\begin{array}{lc}
\displaystyle D^{1a} \cZ = \nabla^{1\alpha} \cZ =0,\qquad D^{2a} \cbZ = \nabla^{2\alpha}\cbZ=0, & \quad (a) \\
\displaystyle \nabla^{2\alpha}D^{2a} \cZ + \nabla^{1\alpha}D^{1a} \cbZ =0 . & \quad (b)
\end{array}
\ee
The constraints (\ref{constr}) leave the following components in the~$N=8$ superfields~$\cZ$,~$\cbZ$:
\be\label{components}
\begin{array}{llll}
z = \cZ, & \bar{z} = \cbZ, & Y^{a\alpha}=D^{2a}\nabla^{2\alpha}\cZ, & \overline{\strut Y}{}^{a\alpha} = -D^{1a}\nabla^{1\alpha}\cbZ=Y^{a\alpha}, \\[0.1em]
\psi^a =D^{2a}\cZ, & \bpsi_a=-D^1_a\cbZ, & \xi^\alpha=\nabla^{2\alpha}\cZ, & \bxi_\alpha=-\nabla^1_\alpha\cbZ,\\[0.1em]
X=-i D^{2a}D^2_a \cZ, & {\bX}=-i D^{1a}D^1_a \cbZ, & Y=-i\nabla^{2\alpha}\nabla^2_\alpha \cZ, &
\bY=-i\nabla^{1\alpha}\nabla^1_\alpha \cbZ,
\end{array}
\ee
where the right hand side of each expression is supposed to be taken upon~$\theta^{ia}=\vartheta^{i\alpha}=0$.
The bosonic auxiliary components~$X$ and~$Y$ are subjected, in virtue of (\ref{constr}), to the additional constraints
\be\label{constr1}
\frac{\partial}{\partial t} \left( X-{\bY}\right)=0,\quad
\frac{\partial}{\partial t} \left( {\bX}- Y\right)=0.
\ee
Simple component counting gives that we have two physical bosons, eight fermions and six auxiliary fields, i.e. {\bf (2,\;8,\;6)} supermultiplet.

When writing down supersymmetric actions, we make use of intuitively understandable notations
$$
F_{AB} \equiv \frac{\partial^2\, \cF(\cZ)}{\partial \cZ{}^A \partial \cZ{}^B}\rule[-0.7em]{0.4pt}{1.9em}\,,\quad
\bF_{AB} \equiv \frac{\partial^2\,\cbF(\cbZ )}{\partial \cbZ{}^A \partial {\cbZ}{}^B} \rule[-0.7em]{0.4pt}{1.9em}\,.
$$
and the following rule for Grassmann integration
$$\int d^2\theta_2 d^2\vartheta_2\equiv \frac{1}{4}\; D^{2a}D^2_a\;\nabla^{2\alpha}\nabla^2_\alpha.$$


\bigskip
\begin{thebibliography}{99}
\bibitem{bh1} P.~Claus, M.~Derix, R.~Kallosh, J.~Kumar, P.K.~Townsend, A.~Van~Proeyen,
Phys.~Rev.~Lett. {\bf 81} (1998) 4553; {\tt  arXiv:hep-th/9804177}.

\bibitem{AFF} V.~de~Alfaro, S.~Fubini, G.~Furlan, Nuovo.~Cimento. {\bf 34A} (1976) 569.

\bibitem{newcfm} S.~Bellucci, E.~Ivanov, S.~Krivonos, Phys.Rev. {\bf D66} (2002)
086001; Erratum-ibid. {\bf D67} (2003) 049901; {\tt
arXiv:hep-th/0206126};\\
S. Bellucci, A. Galajinsky, E. Ivanov, S. Krivonos, Phys.Lett.{\bf B555} (2003) 99;\\
E.~Ivanov, S.~Krivonos, J.~Niederle, Nucl.Phys. {\bf B677} (2004)
485; {\tt arXiv:hep-th/0210196};
A.~Galajinsky, ``Particle dynamics on $AdS_2\, \times S^2$ background with two-form flux'',
{\tt arXiv:0806.1629[hep-th]}.

\bibitem{1} G.W.~Gibbons, Nucl.~Phys. {\bf B207} (1982) 337; \\
P.~Breitenlohner, D.~Maison and G.~Gibbons, Commun.~Math.~Phys. {\bf 120} (1988) 295.

\bibitem{2} S.~Ferrara, G.W.~Gibbons and R.~Kallosh, Nucl.~Phys. {\bf B500} (1997) 75, {\tt arXiv:hep-th/9702103}.

\bibitem{KG} G.~Sierra, P.K.~Townsend, in \emph{``Supersymmetry and Supergravity''}, ed. B.~Milewski
(World Scientific, Singapore, 1983);\\
S.J.~Gates, Nucl. Phys. {\bf B238} (1984) 349.

\bibitem{AM} S.~Ferrara, R.~Kallosh and A.~Strominger, Phys.Rev~{\bf D52} (1995) 5412, {\tt arXiv:hep-th/9508072};\\
A.~Strominger, Phys. Lett.~{\bf B383} (1996) 39, {\tt arXiv:hep-th/9602111};\\
S.~Ferrara and R.~Kallosh, Phys. Rev.~{\bf D54} (1996) 1514, {\tt arXiv:hep-th/9602136};\\
S.~Ferrara and R.~Kallosh, Phys. Rev.~{\bf D54} (1996) 1525, {\tt arXiv:hep-th/9603090};\\
R.~Kallosh, JHEP~{\bf 0512} (2005) 022, {\tt arXiv:hep-th/0510024}.

\bibitem{iv} S.~Bellucci, E.~Ivanov, S.~Krivonos, O.~Lechtenfeld, Nucl. Phys. {\bf B699} (2004) 226,
{\tt arXiv:hep-th/0406015}.

\bibitem{Z1} B.~Zupnik, Nucl. Phys. {\bf B554} (1999) 365, Erratum-ibid.  {\bf B644} (2002) 405;
{\tt arXiv:hep-th/9902038}.

\bibitem{DE} D.-E.~Diaconescu, R.~Entin, Phys. Rev. {\bf D56} (1997) 8045;
{\tt arXiv:hep-th/9706059}.

\bibitem{bkn} S.~Bellucci, S.~Krivonos, A.~Nersessian,  Phys. Lett.  {\bf B605} (2005) 181; {\tt arXiv:hep-th/0410029}.

\bibitem{bks1} S.~Bellucci, S.~Krivonos, A.~Shcherbakov, Phys.Lett. {\bf B612} (2005) 283; {\tt
arXiv:hep-th/0502245};\\
S. Bellucci, S. Krivonos, A. Nersessian, A. Shcherbakov,
``$2k$-dimensional N=8 supersymmetric quantum mechanics'',
Proceedings of the XI International Conference ``Symmetry Methods
in Physics'', 2004, June 21-24, Prague, Czech Republic; {\tt
arXiv:hep-th/0410073}.

\bibitem{leva} E.~Ivanov, S.~Krivonos, V.~Leviant, J. Phys. {\bf A22} (1989) 4201.

\bibitem{ikl1} E.~Ivanov, S.~Krivonos, O.~Lechtenfeld, Class. Quant. Grav. {\bf 21} (2004) 1031-1050;
{\tt  arXiv:hep-th/0310299}.

\bibitem{sorba} L.~Frappat, P.~Sorba, A.Sciarrino, ``Dictionary on Lie superalgebras'',
{\tt arXiv:hep-th/9607161};\\
A.~Van~Proeyen,``Tools for supersymmetry'', {\tt arXiv:hep-th/9910030}.

\bibitem{ikp} E.~Ivanov, S.~Krivonos, A.~Pashnev, Class. Quant. Grav. {\bf 8} (1991) 19.
\end{thebibliography}
\end{document}